\documentclass[twocolumn,aps,tightenlines,floatfix,showpacs]{revtex4}
\usepackage{amsmath}
\usepackage[dvips]{graphicx}
\usepackage{hhline}
\usepackage{epsfig}

\newcommand{\beq}{\begin{equation}}
\newcommand{\eeq}{\end{equation}}
\newcommand{\beqn}{\begin{eqnarray}}
\newcommand{\eeqn}{\end{eqnarray}}

\newcommand{\cJ}{I}

\begin{document}

\title{Neutrinoless double-$\beta$ decay of ${}^{82}$Se in the shell model: beyond closure approximation}

\author{R.A.~Sen'kov$^1$, M.~Horoi$^1$, and B.A.~Brown$^2$}

\affiliation{$^1$ Department of Physics, Central Michigan University, Mount Pleasant, Michigan 48859, USA\\
$^2$ National Superconducting Cyclotron Laboratory and
Department of Physics and Astronomy, Michigan State University, East Lansing, Michigan 48824-1321, USA}


\begin{abstract}
We recently proposed a method 
[R.~A.~Senkov and M.~Horoi, Phys. Rev. C 88, 064312 (2013)] 
to calculate the standard nuclear matrix elements for neutrinoless double-$\beta$ decay ($0\nu\beta\beta$) of ${}^{48}$Ca going beyond the closure approximation. Here we extend this analysis to the important case of ${}^{82}$Se, which was chosen as the base isotope for the upcoming SuperNEMO experiment. We demonstrate that by using a mixed method that considers information from closure and nonclosure approaches, one can get excellent convergence properties for the nuclear matrix elements, which allows one to avoid unmanageable computational costs. We show that in contrast with the closure approximation the mixed approach has a very weak dependence on the average closure energy. The matrix elements for the heavy neutrino-exchange mechanism that could contribute to the $0\nu\beta\beta$ decay of ${}^{82}$Se are also presented.
\end{abstract}

\maketitle

\section{Introduction}

Neutrinoless double-$\beta$ decay ($0\nu\beta\beta$) is one of the most important current topics in physics and provides
unique information on the neutrino properties
\cite{ves12,jermp,tomoda}.
The $0\nu\beta\beta$ decay process and the associated nuclear matrix
elements (NME) were investigated using several approaches including the quasiparticle random phase approximation (QRPA) \cite{ves12}, the
interacting shell model \cite{prl100,prc13}, the interacting boson model \cite{iba-2,iba-prl}, the generator coordinate method \cite{gcm}, and the projected Hartree-Fock Bogoliubov model \cite{phfb}. With the exception of the QRPA \cite{qrpa-J,simvo11},
all other methods use the closure approximation \cite{prc10}.

In this paper we calculate and analyze the nonclosure nuclear matrix elements for the $0\nu\beta\beta$ decay of ${}^{82}$Se, which was chosen 
for the upcoming SuperNEMO experiment \cite{supernemo}.
A standard way to consider a double $\beta$-decay process is 
to present it as a transitional process from an initial nucleus 
to an intermediate nucleus and then to a final nucleus, so that the 
corresponding nuclear matrix elements can be presented as a sum over 
intermediate nuclear states. To calculate these matrix elements
one needs to calculate all the intermediate states, which could 
be a very challenging task, especially for heavy nuclei in realistic 
model spaces. Closure approximation is one possible way to avoid 
unmanageable computational costs. The main idea behind the closure 
approximation is to replace the energies of the intermediate states
with an average energy $\langle E\rangle$, and then the sum
over the intermediate states can be found explicitly 
by using the completeness relation. The uncertainty in the
average closure energy brings an error into the NME, but this error
is not very large (it was estimated to be about 10\% \cite{sh13,prc10}).

With the goal of going beyond the closure approximation but keeping 
a limited number of intermediate states, 
we consider four different approaches: pure closure (we use ``pure closure" to distinguish it from the  ``running closure" approach), running closure, running nonclosure, and mixed approximations. In the running nonclosure approach the exact intermediate energies are used and the sum over intermediate states is performed. Since we cannot calculate all the intermediate states the sum is restricted by a number-of-states cutoff parameter $N$, so that the corresponding matrix elements become functions of the cutoff parameter. The running closure approximation also contains a restricted sum over intermediate states, but with all intermediate energies replaced by the average closure energy $\langle E \rangle$ as it was done for the pure closure approximation. The running closure approximation is introduced to check the convergence properties of the NME and to construct the mixed approximation, which includes information from the nonclosure and closure approaches. In the mixed method
all the intermediate states below the cutoff parameter $N$ are taken into 
account within nonclosure approach, while for the higher states the closure approximation is used.
 
We demonstrate that the mixed-method matrix elements have perfect convergence properties and, at the same time, have very weak dependence on the average closure energy, which allows us to avoid unmanageable computational cost and achieve a high accuracy in NME calculations.
We argue that the mixed approach can be successfully applied to more  computationally challenging cases, such as the $0\nu\beta\beta$ decay of $^{76}$Ge. For the calculations we used a shell-model approach within the realistic $jj44$ model space having the nucleus $^{56}$Ni as a core and the $f_{5/2}, p_{3/2}, p_{1/2},\mbox{ and } g_{9/2}$ orbitals as the valence space. For this model space we used the JUN45 effective interaction \cite{jun45}
fine-tuned for the region of the nuclear chart 
close to $^{82}$Se and $^{76}$Ge.
The nonclosure approach provides information about the contribution
of intermediate states with different spin and parity, which
we obtained for the first time for the light neutrino-exchange mechanisms shell model NME of ${}^{82}$Se. 
It is also interesting to go beyond the closure approximation 
for the NME corresponding to other mechanisms that may contribute to the $0\nu\beta\beta$ decay rates \cite{ves12,prc13,prl13}. Here we extended our approach to the heavy neutrino-exchange mechanisms NME, and we calculated 
for the first time 
the decomposition of this NME vs spin and parity of intermediate states.

The analysis of the $0\nu\beta\beta$ decay NME beyond closure approximation requires knowledge of a large number of one-body transition densities connecting the ground states of the initial and final 
nuclei $^{82}$Se and $^{82}$Kr, respectively, with states of the intermediate nucleus $^{82}$Br. The actual number of intermediate states of $^{82}$Br we have to deal with is $1.0\times10^7$ including all spins and parities. 
As a comparison, for the similar analysis of $^{76}$Ge, one needs to consider about $1.5\times10^8$ states in the intermediate nucleus $^{76}$As. However, we demonstrate that using the mixed-method approximation, with only a few hundred intermediate states of each $J^\pi$, it is possible to obtain an accurate value of the nuclear matrix element.

The paper is organized as follows. Section II gives a brief description
of the $0\nu\beta\beta$ NME relevant for the
distinction between the pure closure, running closure, running nonclosure, and mixed approximations. In Sec. III we analyze the numerical results, and Sec. IV is devoted to conclusions and outlook.

\section{The nuclear matrix element}

In this section we briefly review the method developed in Ref. \cite{sh13} for calculating the beyond closure NME for $0\nu\beta\beta$ decay. We start with the light neutrino-exchange mechanism of a $0\nu\beta\beta$ decay. 
The corresponding decay rate can be written as \cite{ves12}
\beq \label{nme0}
\left[ T^{0\nu}_{1/2} \right]^{-1} = G^{0\nu} | M^{0\nu} |^2
\left(\frac{\langle m_{\beta \beta}\rangle}{m_e}\right )^2,
\eeq
where $G^{0\nu}$ is the known phase-space factor  \cite{kipf12},
$M^{0\nu}$ is the nuclear matrix element, and $ \langle m_{\beta \beta}\rangle$ is the effective neutrino mass defined by the neutrino
mass eigenvalues $m_k$ and the elements of neutrino mixing matrix
$U_{ek}$ \cite{ves12},
\beq
\langle m_{\beta \beta}\rangle = \left| \sum_k m_k U^2_{ek} \right|.
\eeq

The nuclear matrix element  $M^{0\nu}$ can be presented as a sum
of Gamow-Teller ($GT$), Fermi ($F$), and Tensor ($T$)
matrix elements (see, for example, Refs. \cite{sh13,prc10}),
\beq \label{nme1}
M^{0\nu} = M^{0\nu}_{GT} - \left( \frac{g_{V}}{g_{A}} \right)^2
M^{0\nu}_{F} + M^{0\nu}_{T},
\eeq
where $g_{V}$ and $g_{A}$ are the vector and axial constants,
correspondingly. In our calculations we use $g_{V}=1$ and $g_{A}=1.254$.

The matrix elements in Eq. (\ref{nme1}) describe the transition from the
initial state $|i\rangle$ of 
$^{82}$Se to the final state $|f\rangle$ of 
$^{82}$Kr.
They can be presented as sums over
intermediate states $| \kappa \rangle=|E_\kappa, J^\pi_\kappa \rangle$
of 
$^{82}$Br,
\beq \label{nme2}
M^{0\nu}_{\alpha}=\sum_{\kappa} \sum_{1234} \langle 1 3 | {\cal O}_{\alpha} | 2 4\rangle
\langle f |  \hat{c}^\dagger_{3} {\hat{c}}_4 | \kappa \rangle
\langle \kappa |  \hat{c}^\dagger_{1} {\hat{c}}_2 | i \rangle.
\eeq
Here $\alpha=\{ GT, F, T \}$, the operators ${\cal O}_{\alpha}$ contain
neutrino potentials with spin and isospin dependence, and they explicitly depend on the energy of intermediate states $|\kappa\rangle$:
${\cal O}_{\alpha}={\cal O}_{\alpha}(E_\kappa)$.
The full expression for these operators and the calculation
details for two-body matrix elements $\langle 1 3 | {\cal O}_{\alpha} | 2 4\rangle$ can be found in Ref. \cite{sh13}. 
Equation (\ref{nme2}) presents exact or nonclosure NME, which we are going to analyze using four different approximations mentioned in the introduction.  

In the {\it pure closure approximation}, one needs to replace the energies of intermediate states $|\kappa\rangle$ in the operators ${\cal O}_{\alpha}(E_\kappa)$ by a constant value $\langle E\rangle$ (we call it average closure energy or average energy)
\beq
{\cal O}_\alpha \rightarrow \tilde{{\cal O}}_\alpha \equiv {\cal O}_\alpha(\langle E \rangle).
\eeq
Thus the sum over intermediate states in the matrix elements 
(\ref{nme2}) can be found explicitly by using the completeness relation.
The pure closure NME can be presented as (in this equation and below
the sum over repeated indices $\{1,2,3,4\}$ is assumed and will be omitted):
\beq \label{nme3}
{\cal M}^{0\nu}_{\alpha}=
\langle 1 3 | \tilde{{\cal O}}_{\alpha} | 2 4\rangle
\langle f |  \hat{c}^\dagger_{3} {{c}}_4 \hat{c}^\dagger_{1} {\hat{c}}_2
| i \rangle.
\eeq

For the {\it nonclosure approach} one needs to calculate the
sum in Eq. (\ref{nme2}) completely, which could be
challenging due to the large number of intermediate states
$|\kappa \rangle$. Since all intermediate states cannot be included, we
introduce a number-of-state cutoff parameter $N$ and the corresponding
 running nonclosure NME, which are represented as
\beq\label{nme4}
M^{0\nu}_{\alpha}(N)=\sum_{\kappa \leq N}
\langle 1 3 | {\cal O}_{\alpha} | 2 4\rangle
\langle f |  \hat{c}^\dagger_{3} {\hat{c}}_4 | \kappa \rangle
\langle \kappa |  \hat{c}^\dagger_{1} {\hat{c}}_2 | i \rangle,
\eeq
where we take into account only states with $\kappa \leq N$.
In the limit of the large cutoff parameter $N$, the running 
nonclosure matrix element $M^{0\nu}_{\alpha}(N)$ approaches its exact 
nonclosure value (\ref{nme2}).

In the mixed approximation we calculate the running nonclosure 
matrix elements for the states with $\kappa \leq N$ and keep $N$ as large as 
possible to perform the shell-model computation. For the higher states with $\kappa>N$, we use the closure approximation. To do so we introduce first
the {\it running closure approximation}:
\beq\label{nme5}
{\cal M}^{0\nu}_{\alpha}(N)=\sum_{\kappa \leq N}
\langle 1 3 | \tilde{{\cal O}}_{\alpha} | 2 4\rangle
\langle f |  \hat{c}^\dagger_{3} {\hat{c}}_4 | \kappa \rangle
\langle \kappa |  \hat{c}^\dagger_{1} {\hat{c}}_2 | i \rangle.
\eeq
The difference between Eqs. (\ref{nme4}) and (\ref{nme5}) is that for
the running nonclosure approach the operators ${\cal O}_\alpha$
are functions of the excitation energy $E_\kappa$, while for the running closure approximation the same operators $\tilde{{\cal O}}_\alpha$ are functions of the average closure  energy $\langle E \rangle$.
In the limit of the large cutoff parameter $N$, the running closure matrix elements approach the pure closure limit (\ref{nme3}): $ {\cal M}^{0\nu}_{\alpha}(N) \rightarrow {\cal M}^{0\nu}_{\alpha}$.

The {\it mixed-method} matrix elements are defined as 
\cite{sh13}
\beq\label{nme6}
{\bar M^{0\nu}_{\alpha}}(N)=
{M}^{0\nu}_{\alpha}(N)-{\cal M}^{0\nu}_{\alpha}(N)
+{\cal M}^{0\nu}_{\alpha}.
\eeq
We expect that the mixed NME converges significantly faster than 
running nonclosure and closure matrix elements separately. The mixed NME start with the pure closure values (\ref{nme3}) at $N=0$ and reach the nonclosure values (\ref{nme2}) at $N\rightarrow\infty$ (see solid and dotted lines in Fig. \ref{fig3}). It is also expected that the mixed NME will have much weaker dependence on the average energy $\langle E \rangle$ compared with the pure and running closure NME.

\begin{figure}
\begin{center}
\includegraphics[width=0.47\textwidth]{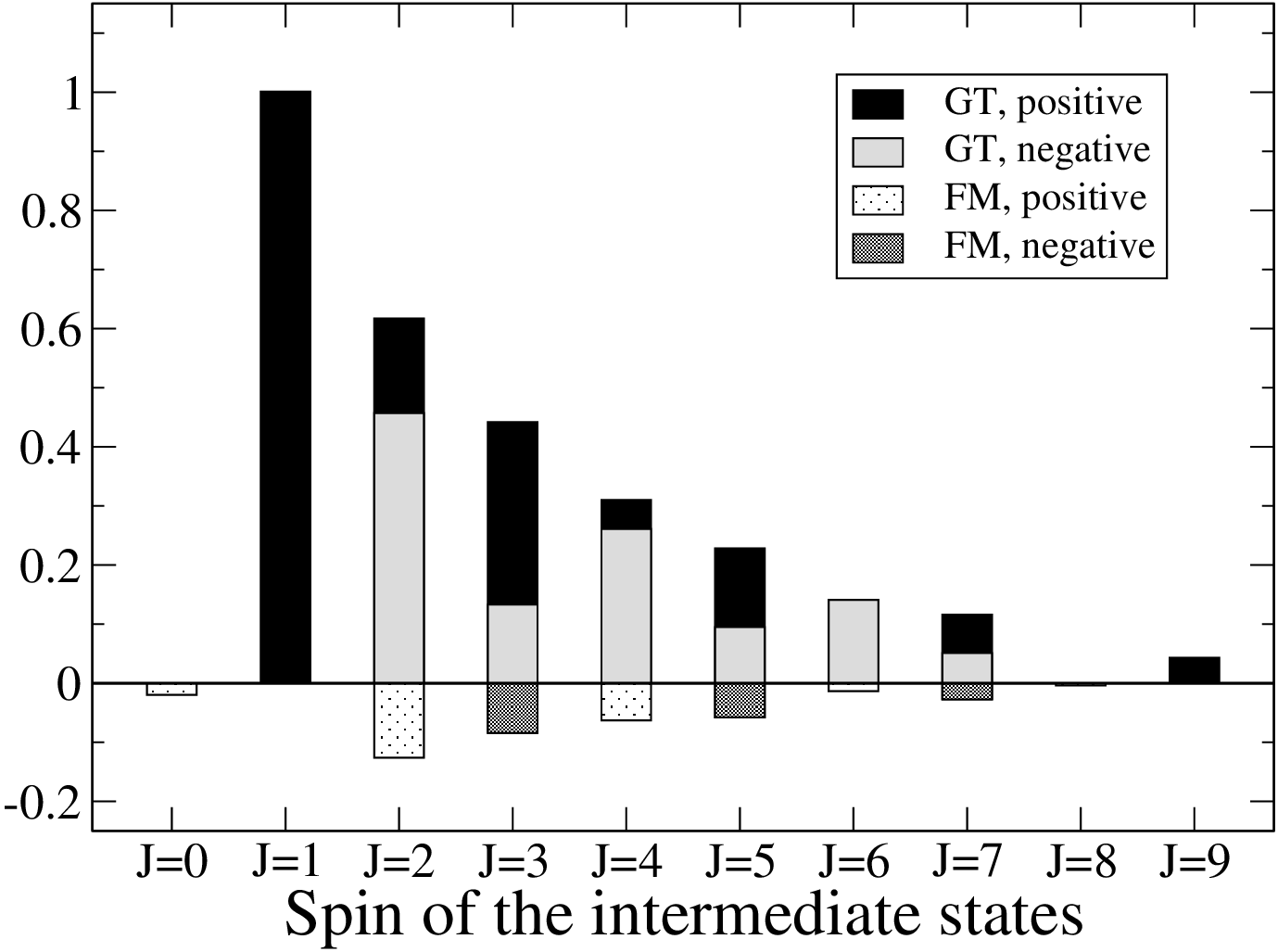}
\caption{$J_\kappa$ decomposition: contributions of the intermediate states $|\kappa \rangle$ with certain spin and parity $J^\pi$ to the running nonclosure Gamow-Teller (solid colors) and Fermi (dashed colors) matrix elements for the $0\nu\beta\beta$ decay of ${}^{82}$Se (light neutrino exchange). Solid black and dashed white bars correspond to the positive-parity states, while solid gray and shaded black bars represent the states with negative parity. CD-Bonn SRC parametrization was used.\\}\label{fig1}
\end{center}
\end{figure}

The heavy neutrino-exchange matrix elements for a $0\nu\beta\beta$ decay process are defined similarly to Eqs. (\ref{nme0}) and (\ref{nme1}), and the corresponding contribution to the total decay rate can be written as
\beq\label{hnme0}
\left[ T^{0\nu}_{1/2} \right]_{\mbox{heavy}}^{-1} = G^{0\nu} | M^{0\nu}_N |^2
|\eta_{NR}|^2,
\eeq
where the heavy neutrino-exchange matrix elements $M^{0\nu}_N$ have the 
structure similar to the light neutrino-exchange NME (\ref{nme1}) and  (\ref{nme2}), while the parameter $\eta_{NR}$ depends on the heavy neutrino masses (for more details see, for example, Ref. \cite{prc13}).
The difference between the heavy and light neutrino-exchange
mechanisms is that the heavy neutrino-exchange NME does not depend on the energy of intermediate states. The standard perturbation theory energy denominator is reduced to the heavy neutrino mass and all other energies can be neglected. Thus for the heavy neutrino-exchange mechanism the pure closure
approach provides the exact nonclosure matrix element. 

\section{Results}

\begin{figure}
\begin{center}
\includegraphics[width=0.47\textwidth]{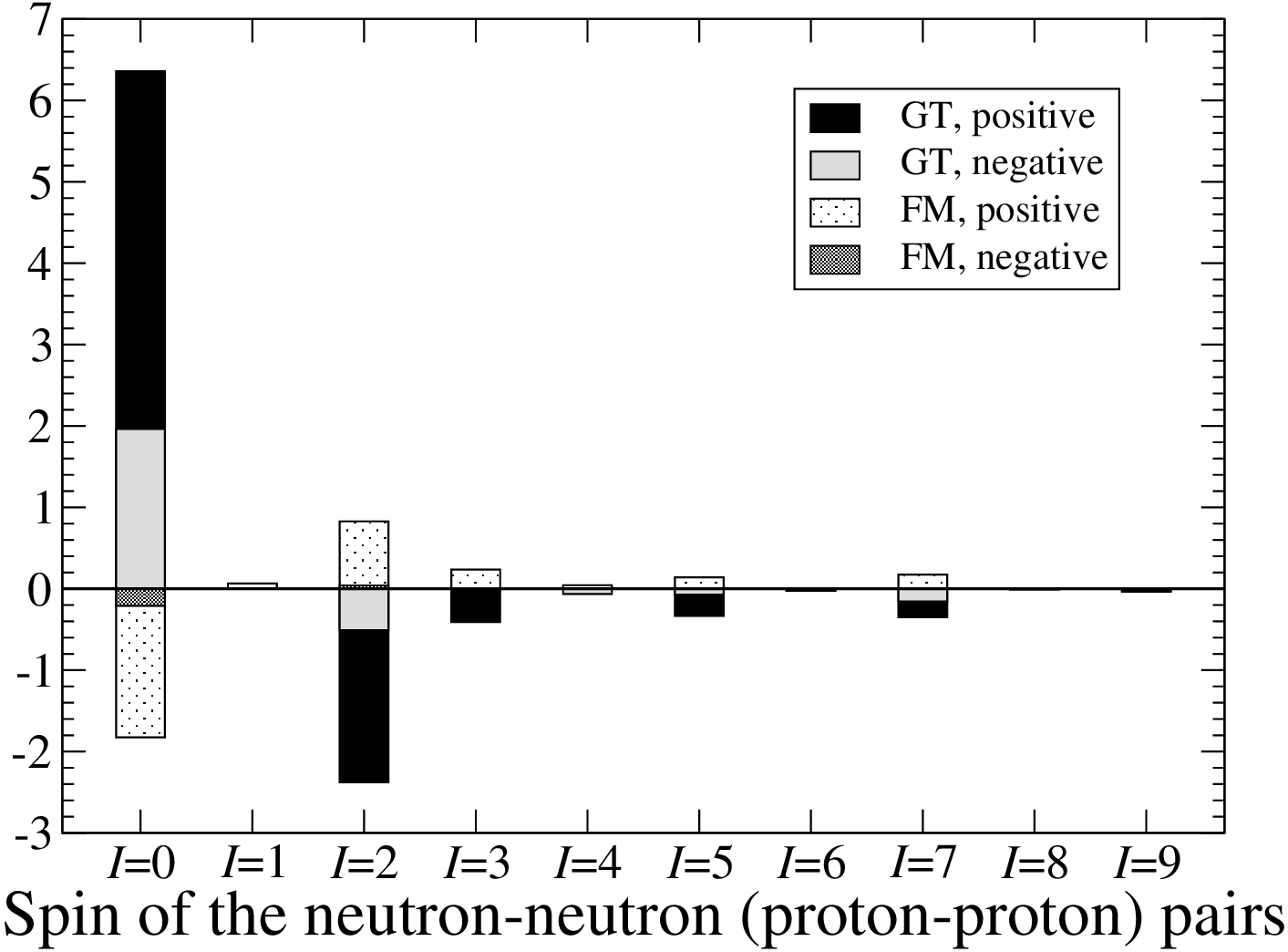}
\caption{$\cJ$ decomposition: contributions to the running nonclosure Gamow-Teller and Fermi matrix elements for the $0\nu\beta\beta$ decay of ${}^{82}$Se (light neutrino exchange) from the configurations when two initial neutrons $|24\rangle$ (and two final protons $|13\rangle$) have certain total spin $\cJ$, $\langle 13, \cJ| {\cal O}^\alpha | 24, \cJ \rangle$.
The grayscale pattern and the SRC parametrization scheme are the same as in Fig. \ref{fig1}.\\}\label{fig2}
\end{center}
\end{figure}

Figure \ref{fig1} presents the Gamow-Tellor and the Fermi (multiplied by the factor $(g_V/g_A)^2$) running nonclosure light neutrino-exchange matrix elements calculated for the fixed spin and parity $J_\kappa^\pi$ of intermediate states $|\kappa\rangle$ \cite{sh13}. Knowing this $J_\kappa$ decomposition one can easily find the total matrix elements as a sum over all the spin contributions: ${M}_\alpha=\sum_{J_\kappa} {M}_\alpha(J_\kappa)$. The Gamow-Teller matrix elements are all positive (presented with solid colors) and the Fermi matrix elements are all negative (presented with dashed colors). Since contributions of the Tensor NME are negligibly small (see Table \ref{tbl1}) the total size of each bar in Fig. \ref{fig1} roughly corresponds to the total NME for given $J_\kappa$. The model space used is large enough to allow contributions from both the negative parity (presented with solid gray and dashed black bars) and the positive parity (presented with solid black and dashed white bars) intermediate states, but $jj44$ model space is still imperfect as it misses the $f_{7/2}$ orbital (the spin-orbit partner of the $f_{5/2}$ orbital) and the $g_{7/2}$ orbital (the spin-orbit partner of the $g_{9/2}$ orbital). As the result, the Ikeda sum rule is not satisfied and 
some contributions, such as ${M}^{0\nu}_{GT}(J_\kappa^\pi=6^+,8^+)$ and ${M}^{0\nu}_{FM}(J_\kappa^\pi=1^-)$, are missing. This deficiency is reflected in the NME for the two-neutrino double-$\beta$ decay of $^{82}$Se, which is about twice its experimental value of 0.1 MeV$^{-1}$ \cite{barabash10} when one uses the standard quenching factor of 0.74 for the Gamow-Teller operator. One can get the experimental value of this particular NME by decreasing the quenching factor to about 0.54 (see also Table 2 of Ref. \cite{2njun45}). The situation, however, is not as dramatic as is in the case of $^{136}$Xe \cite{prl13}, which required the consideration of the missing spin-orbit partner orbitals. Unfortunately, we cannot consider the complete model space, such as $pfg$, due to its unmanageably large dimensions.

The $J$ decomposition for the $0\nu\beta\beta$ decay of $^{82}$Se is presented for the first time 
here as a result of the shell-model analysis. The one-body transition densities
$\langle f |  \hat{c}^\dagger_{3} {\hat{c}}_4 | \kappa \rangle$
and
$\langle \kappa |  \hat{c}^\dagger_{1} {\hat{c}}_2 | i \rangle$)
were calculated for the first 250 intermediate states $|\kappa\rangle$ for each $J_\kappa^\pi$ with the NUSHELLX code \cite{nushellx} at the MSU High Performance Computer Center \cite{hpcc}.
We used the JUN45 two-body interaction \cite{jun45} in $jj44$ model space.
In the calculations we included the short-range correlations (SRC) parametrization based on the CD-Bonn potential and the standard
nucleon finite-size effects \cite{prc10}. The other parameters of the calculation are the ground-state energies and $Q$ value,
$\left(E_{g.s.}({}^{82}\mbox{Br})-E_{g.s.}({}^{82}\mbox{Se})+Q_{\beta\beta}/2\right)=1.595\,\mbox{MeV}$; the oscillator length, $b_{osc}=2.143\,\mbox{fm}$; and the nuclear radius, $R_0=5.213\,\mbox{fm}$.

\begin{figure}
\begin{center}
\includegraphics[width=0.47\textwidth]{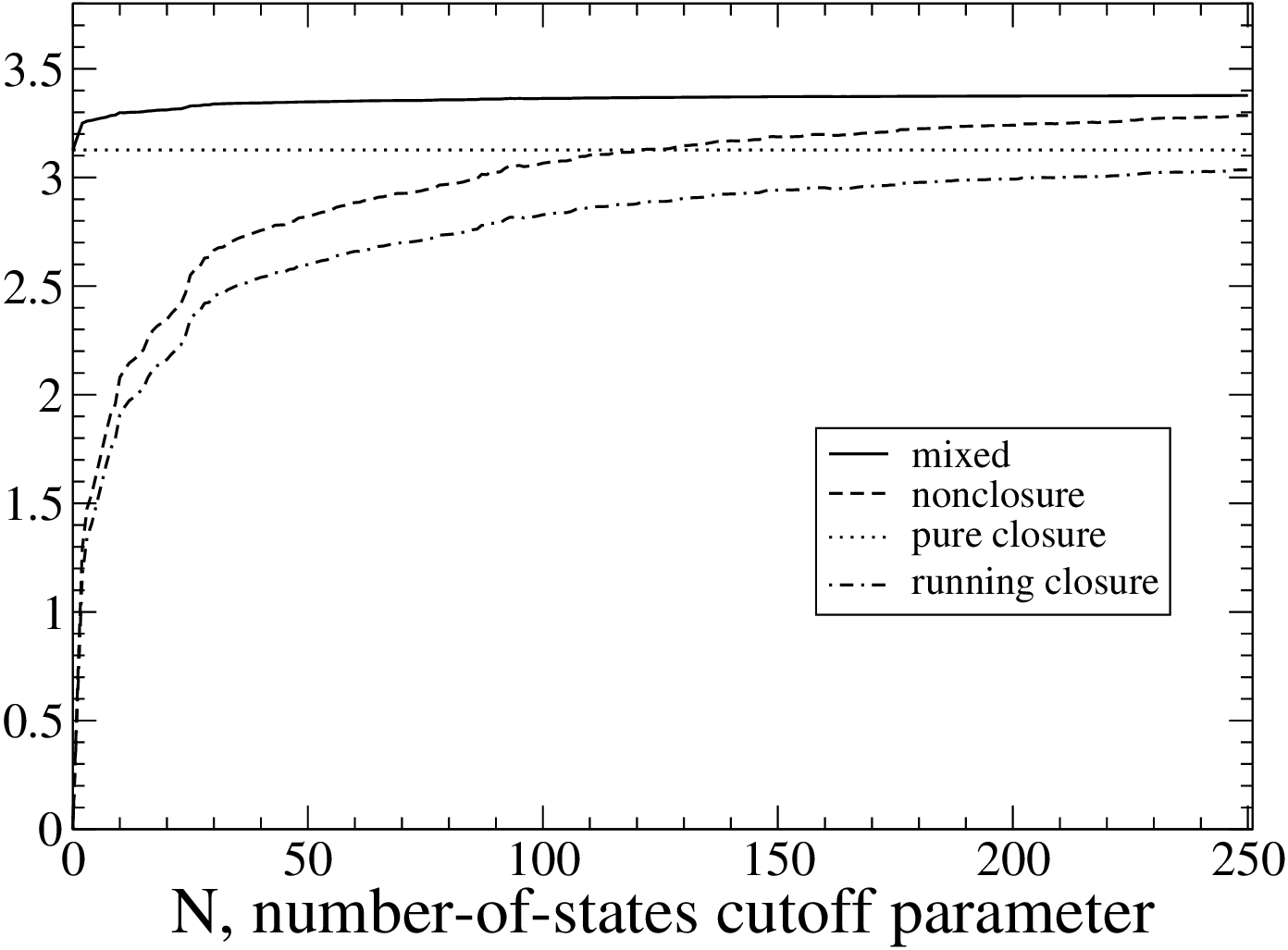}
\caption{Convergence of NME for the $0\nu\beta\beta$ decay of ${}^{82}$Se (light neutrino exchange) calculated within different approximations: mixed (solid curve), running nonclosure  (dashed curve), pure closure (dotted curve, does not depend on $N$), and running closure (dash-dotted curve). All calculations were done with CD-Bonn SRC and $\langle E \rangle=10$ MeV.\\}\label{fig3}
\end{center}
\end{figure}

\begin{figure}
\begin{center}
\includegraphics[width=0.47\textwidth]{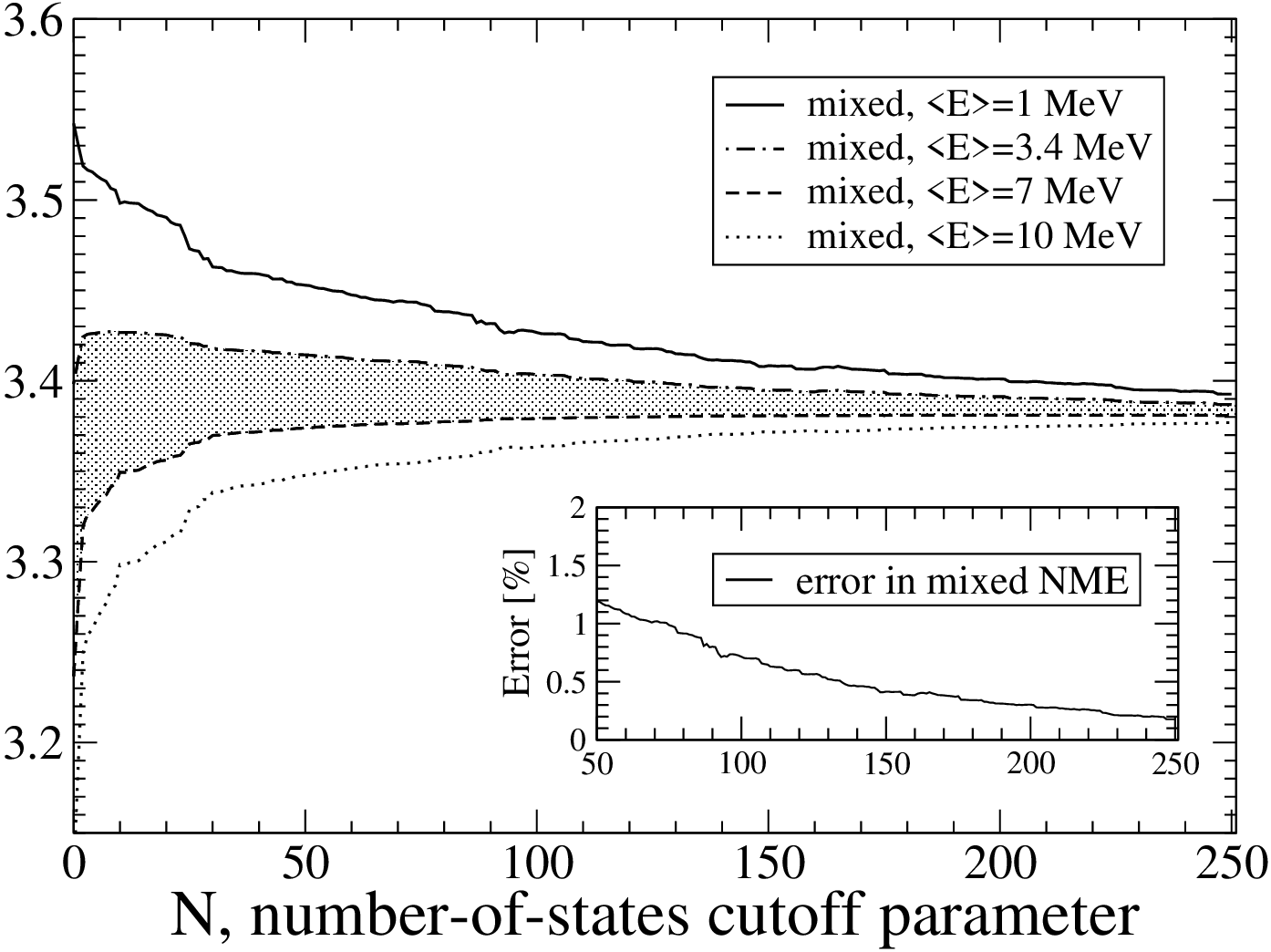}
\caption{Dependence of mixed NME (light neutrino exchange)
on the cutoff parameter $N$ calculated for different average closure energies $\langle E \rangle$. The main panel: $\langle E \rangle=1$ MeV (solid curve), $\langle E \rangle=3.4$ MeV  (dash-dotted curve), $\langle E \rangle=7$ MeV (dashed curve), and $\langle E \rangle=1$ MeV (dotted curve).  The insert shows the uncertainty in the value of mixed NME corresponding to the shaded area from the main panel.\\}\label{fig4}
\end{center}
\end{figure}

Figure \ref{fig2} presents another possible way to decouple the NME of $0\nu\beta\beta$ decay process. In this decoupling scheme we consider two-body matrix elements $\langle 1 3 | {\cal O}_\alpha | 2 4 \rangle$ with
the single-particle states $|1\rangle$ and $|3\rangle$ (proton states)
and the states $|2\rangle$ and $|4\rangle$ (neutron states) have been coupled
to certain common spin $\cJ$, so that the total NME can be presented as $M_\alpha = \sum_{\cJ} M_\alpha(\cJ)$.
The details of such decoupling can be found in Ref. \cite{sh13}.
The grayscale scheme in Fig. \ref{fig2} is similar to the scheme used in
Fig. \ref{fig1}. In contrast to the intermediate spin decoupling, where all the spins $J_\kappa$ contribute coherently,
in the $\cJ$-decoupling scheme we see a significant cancellation between 
$\cJ=0$ and $\cJ=2$.
Similar effects have been observed in shell-model analysis \cite{sh13} and in seniority-truncation studies \cite{menendez-sen} of the NME of $^{48}$Ca (see also Ref. \cite{npa818} for effects of higher seniority in shell-model calculations). QRPA results are available for
heavier nuclei (see, e.g., Fig. 1 of Ref. \cite{qrpa-J}), for which the
$\cJ=0$ and $\cJ=2$ contributions are still dominant, but the cancellation effect is significantly reduced.

\begin{figure}
\begin{center}
\includegraphics[width=0.47\textwidth]{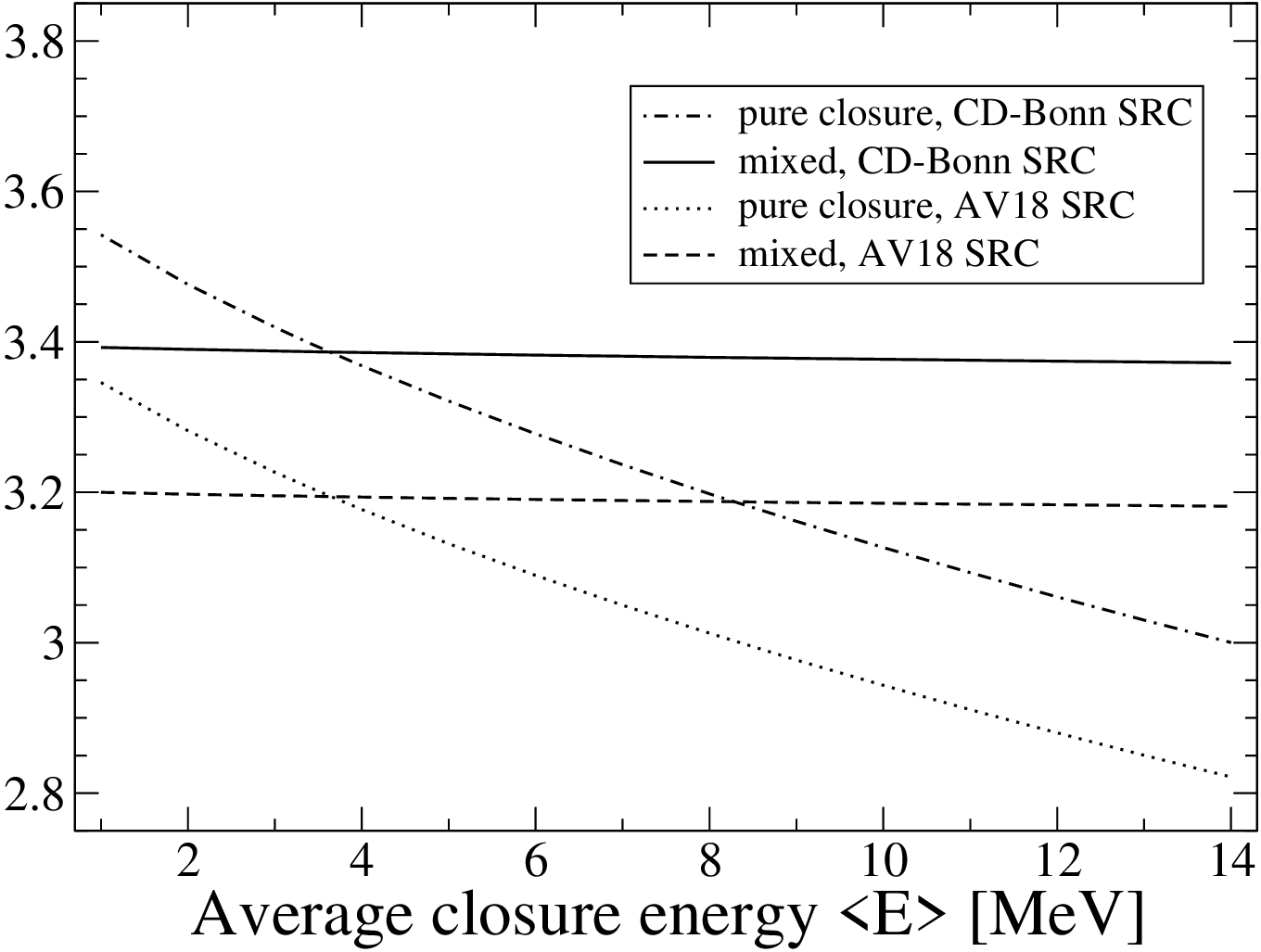}
\caption{Dependence of mixed and pure closure total NME for the $0\nu\beta\beta$ decay of ${}^{82}$Se (light neutrino exchange) on the average closure energy $\langle E \rangle$. Matrix elements presented as: mixed with CD-Bonn SRC  (solid curve),  closure with CD-Bonn SRC (dash-dotted), mixed with AV18 SRC (dashed curve), and closure with AV18 (dotted curve).\\}\label{fig5}
\end{center}
\end{figure}

Figure \ref{fig3} shows the convergence of the total nuclear matrix elements for the light neutrino-exchange $0\nu\beta\beta$ decay of $^{82}$Se calculated within different approximations. The solid line represents the mixed matrix element defined by Eq. (\ref{nme6}), the running nonclosure total matrix element Eq. (\ref{nme4}) is presented by the dashed line, the pure closure approximation defined by Eq. (\ref{nme3}) is presented by the dotted line, and finally the running closure matrix element Eq. (\ref{nme5}) is presented by the dash-dotted line. One can see that even if one can include up to 250 intermediate states the running closure matrix element is still about 4\% smaller than the pure closure limit. The running closure and the running nonclosure NME do not converge fast enough to provide a good calculation accuracy. To improve the accuracy we need to include more intermediate states (which is already hard to calculate for $^{82}$Se and practically impossible for the $0\nu\beta\beta$ decay of $^{76}$Ge) or we can use the mixed approximation, which has much better convergence properties: The solid line in Fig. \ref{fig3} becomes flat already after the first 30--50 states.
Figure \ref{fig4} presents convergence properties of the mixed NME in a more enhanced form.

Figure \ref{fig4} allows us to estimate the uncertainties associated with the mixed approximation; it contains only mixed NME calculated for different average closure energies: $\langle E\rangle=1$ MeV (solid line), $\langle E\rangle=3.4$ MeV (dash-dotted line), $\langle E\rangle=7$ MeV (dashed line), and $\langle E\rangle=10$ MeV (dotted line). Our lack of knowledge of the average energy defines the calculation accuracy. If we restrict the range for average energy to 3.4--7.0 MeV (which is quite reasonable since one curve approaches the final NME from above and the other approaches it from  below, so the true NME should be confined somewhere in between), then the uncertainty in the mixed NME can be presented by the shaded area in the main panel. The insert in Figure \ref{fig4} presents the error in the mixed NME associated with the shaded area from the main panel. One can see that the mixed approximation provides an accuracy of less than 1\% for only 50--100 first intermediate states for each $J^\pi$. It can also be seen that there is no need to increase the number of intermediate states; 250 states are more than enough to obtain very good accuracy.

\begin{table}[ht]
\begin{center}
\begin{tabular}{||l|c|c|c|c||}
\hhline{|t:=:=:=:=:=:t|}
\hhline{||-----||}
\rule{0cm}{0.33cm} & Pure closure  & Run. closure & Run. nonclosure & \; Mixed \; \\
\hhline{||-----||}
\rule{0cm}{0.33cm} $M^{0\nu}_{GT}$ & 2.750 & 2.664 & 2.898 & 2.983 \\
\hhline{||-----||}
\rule{0cm}{0.33cm}  $M^{0\nu}_F$ & -0.607 & -0.594 & -0.620 & -0.632 \\
\hhline{||-----||}
\rule{0cm}{0.33cm}  $M^{0\nu}_T$ & -0.011 & -0.008 & -0.007 & -0.011 \\
\hhline{||-----||}
\rule{0cm}{0.33cm}  $M^{0\nu}_{tot}$ & 3.127 & 3.035 & 3.285 & 3.377 \\
\hhline{|b:=:=:=:=:=:b|}
\end{tabular}
\caption{NME for the $0\nu\beta\beta$ decay of ${}^{82}$Se  (light neutrino exchange) calculated within different approximations. All calculations were done with CD-Bonn SRC parametrization and for the average closure energy $\langle E \rangle=10.08$ MeV. The difference between mixed and pure closure total NME is about 8\%.\\}\label{tbl1}
\end{center}
\end{table}

\begin{table}[ht]
\begin{center}
\begin{tabular}{||l|c|c|c|c||}
\hhline{|t:=:=:=:=:=:t|}
\hhline{||-----||}
\rule{0cm}{0.33cm} SRC  & $M^{0\nu}_{GT}$ & $M^{0\nu}_F$ & $M^{0\nu}_T$ & $M^{0\nu}_{total}$\\
\hhline{||-----||}
\rule{0cm}{0.33cm} None & \; 2.898 \; & \; -0.590 \; & \; -0.011 \; & \; 3.262 \; \\
\hhline{||-----||}
\rule{0cm}{0.33cm} Miller-Spencer &  2.337 & -0.419 & -0.011 & 2.593\\
\hhline{||-----||}
\rule{0cm}{0.33cm} CD-Bonn & 2.993 & -0.633 & -0.011 & 3.387 \\
\hhline{||-----||}
\rule{0cm}{0.33cm} AV18 & 2.831 & -0.585 & -0.011 & 3.195 \\
\hhline{|b:=:=:=:=:=:b|}
\end{tabular}
\caption{Mixed NME for the $0\nu\beta\beta$ decay of ${}^{82}$Se
(light neutrino exchange) calculated with different SRC parametrization schemes \cite{prc10}, $\langle E\rangle=3.4\,$MeV.\\
\\}\label{tbl2}
\end{center}
\end{table}

Figure \ref{fig5} shows the difference in the average energy dependence between the mixed and pure closure NME. The average energy varies from 1 to 14 MeV. The solid and dashed lines present the mixed NME calculated with CD-Bonn and AV18 SRC \cite{prc10} correspondingly; these matrix elements have a very weak dependence on average energy. The dash-dotted and dotted lines present the pure closure matrix elements which have stronger dependence on $\langle E\rangle$. While the closure NME varies by 15\%, which is consistent with the similar 
estimates for $^{48}$Ca \cite{sh13,prc10}, the mixed NME only varies by 0.6\%. If we choose the average energy close to 3.4 MeV we can reproduce the mixed results in the framework of the closure approximation.

\begin{figure}
\begin{center}
\includegraphics[width=0.47\textwidth]{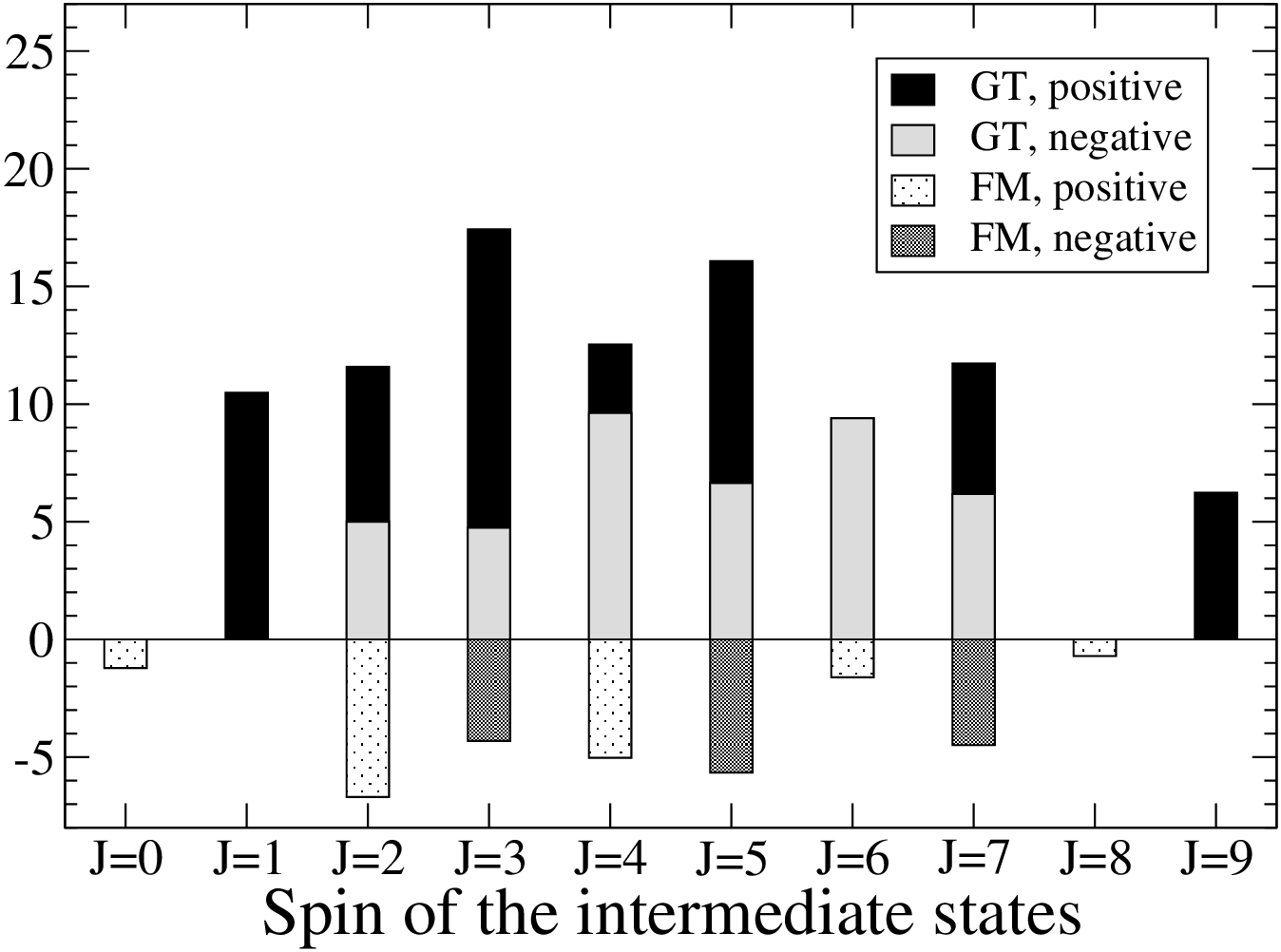}
\caption{$J_\kappa$ decomposition: contributions of the intermediate states $|\kappa \rangle$ with certain spin and parity $J^\pi$ to the Gamow-Teller (solid colors) and Fermi (dashed colors) matrix elements for the $0\nu\beta\beta$ decay of ${}^{82}$Se (heavy neutrino exchange). Solid black and dashed white bars correspond to the positive-parity states, while solid gray and dashed black bars represent the states with negative parity. All calculations were done with CD-Bonn SRC.\\}\label{fig6}
\end{center}
\end{figure}

\begin{table*}[t]
\begin{center}
\begin{tabular}{||ll|c|c|c|c|c|c|c||}
\hhline{|t:=:=:=:=:=:=:=:=:=:t|}
\hhline{||---------||}
\rule{0cm}{0.33cm} 
& & ISM & ISM & QRPA(TBC) & RQRPA(TBC) & QRPA(J) & IBM-2 & EDF \\
\rule{0cm}{0.33cm} 
& SRC & present & \cite{npa818} & \cite{npa793,qrpa-tbc} & \cite{npa793,qrpa-tbc} &
\cite{suh07} & \cite{iba-2,iba-prl} & \cite{gcm} \\
\hhline{||---------||}
\rule{0cm}{0.33cm} $M_{total}^{0\nu}$, & \mbox{Miller-Spencer} & 
2.59 & 2.18 & 4.02 & 3.49 & 2.77 & 4.41 & \\
\hhline{~--------||}
\rule{0cm}{0.33cm} & \mbox{CD-Bonn} & 
3.39 &      & 5.65 & 4.86 &      &      &  \\
\hhline{~--------||}
\rule{0cm}{0.33cm} & \mbox{AV18} & 
3.20 &      & 5.19 & 4.44 &      & 4.84 &  \\
\hhline{~--------||}
\rule{0cm}{0.33cm} & \mbox{UCOM} & 
     & 2.64 &      &      & 3.72 &      & 4.22 \\
\hhline{|b:=:=:=:=:=:=:=:=:=:b|}
\end{tabular}
\parbox{13.2cm}{\caption{Comparison of the total NME for the $0\nu\beta\beta$ decay of ${}^{82}$Se (light neutrino exchange) calculated with
different approaches and with different SRC parametrizations schemes. 
$g_A=1.254$ is used for the axial-vector 
coupling constant, except for IBM-2, which uses $g_A=1.269$ \cite{iba-prl}.\\}\label{tbl4}}
\end{center}
\end{table*}

\begin{table}[ht]
\begin{center}
\begin{tabular}{||l|c|c|c|c||}
\hhline{|t:=:=:=:=:=:t|}
\hhline{||-----||}
\rule{0cm}{0.33cm} SRC, approximation  & $M^{0\nu}_{GT}$ & $M^{0\nu}_F$ & $M^{0\nu}_T$ & $M^{0\nu}_{total}$\\
\hhline{||-----||}
\rule{0cm}{0.33cm} CD-Bonn, pure closure & \, 150 \, & \, -58.1 \, & \, -0.55 \, & \, 187 \, \\
\hhline{||-----||}
\rule{0cm}{0.33cm} CD-Bonn, run.~closure & 149 & -56.9 & -0.83 & 184 \\
\hhline{||-----||}
\rule{0cm}{0.33cm} AV18, pure closure & 99.2 & -48.4 & -0.55 & 129 \\
\hhline{||-----||}
\rule{0cm}{0.33cm} AV18, run.~closure & 97.3 & -47.4 & -0.83 & 127 \\
\hhline{|b:=:=:=:=:=:b|}
\end{tabular}
\caption{Closure NME for the $0\nu\beta\beta$ decay of ${}^{82}$Se
(heavy neutrino exchange) calculated with different SRC parametrizations schemes \cite{prc10}.\\}\label{tbl3}
\end{center}
\end{table}

Table \ref{tbl1} summarizes the differences in the light neutrino-exchange NME calculated within different approximations. The NME for the $0\nu\beta\beta$ decay of $^{82}$Se predicted by the mixed approach is about 8\% greater than the NME obtained with closure approximation when calculated with the average energy $\langle E\rangle=10.08$ MeV, often used in the literature \cite{tomoda,iba-2}. Similar results for other isotopes were reported in Fig. 4 of Ref. \cite{simvo11} obtained within the QRPA approximation. Table \ref{tbl2} presents the mixed $^{82}$Se NME calculations performed with different SRC parametrization sets from Ref. \cite{prc10}.

Table \ref{tbl4} presents a comparison of our results with the recent calculations of $0\nu\beta\beta$ decay of $^{82}$Se (light neutrino exchange). There are five different approaches for the calculation of $0\nu\beta\beta$ decay NME presented in the table: interacting shell model approach (ISM) \cite{npa818}; quasiparticle random phase approximation, T{\"u}ebingen-Bratislava-Caltech group [(R)QRPA(TBC)] \cite{npa793,qrpa-tbc}; quasiparticle random phase approximation, Jyv{\"a}skyl{\"a} group [QRPA(J)] \cite{suh07}; interacting boson model (IBM-2) \cite{iba-2,iba-prl}; and generator coordinate method (EDF) \cite{gcm}. The value $g_A=1.254$ is used in most of the calculations, except for IBM-2, which uses the axial-vector coupling constant $g_A=1.269$ \cite{iba-prl}. 
It is useful to note that the difference between Miller-Spencer and UCOM matrix elements can be accounted for by multiplying the UCOM matrix elements by about 0.8 \cite{ves12,qrpa-J}. 

Figure \ref{fig6} and Table \ref{tbl3} summarize the results for the heavy-neutrino exchange $0\nu\beta\beta$ decay of $^{82}$Se. By comparing Figs. \ref{fig1} and \ref{fig6} one can see that the heavy neutrino-exchange NME do not vanish with the large intermediate spins $J_\kappa$. The heavy-neutrino potentials have a strong short-range part, so the contributions from the large neutrino momentum, which are responsible for the higher spin contributions, are not suppressed. These results are also important because 
for the first time
 a decomposition of this shell model matrix element in spins and parities of 
intermediate states is reported, and it could be used for comparison with  
results of other methods, such as QRPA.

\section{Conclusions and Outlook}

In conclusion, we investigated the beyond closure NME
for the $0\nu\beta\beta$  decay of
$^{82}$Se using 
for the first time 
 shell-model techniques in the realistic $jj44$ shell valence space and the fine-tuned JUN45 effective interaction.

We demonstrated that the mixed-method NME converge very quickly and that by 
using only a few hundred intermediate states we can achieve
high computational accuracy. As in the case of ${}^{48}$Ca \cite{sh13},
we obtained an increase of about 8\% of the nonclosure NME 
compared to the closure result calculated with the standard average energy.
Therefore, for the $jj44$ model space, the JUN45 effective interaction, and the more realistic CD-Bonn- and AV18-based SRC, we
predict a light neutrino-exchange shell model NME for the $0\nu\beta\beta$  decay of $^{82}$Se in the range 
\beq\label{eq11}
M^{0\nu} = 3.3 \pm 0.1,
\eeq
where the error was estimated based on the NME calculated with different SRC parametrization sets (see Table \ref{tbl2}), while the uncertainty associated with the average energy $\langle E\rangle$ is negligible (see Fig. \ref{fig4}). 
For direct comparison with the majority of previous calculations (see, e.g.,  Table \ref{tbl4}) all our calculations were performed with the $g_A=1.254$, while a most recent value for the axial-vector coupling constant is $g_A=1.269$. Changing the axial-vector constant to its recent value decreases the average NME (\ref{eq11}) by only 0.5\%, which can be neglected. 

For the same model space, effective interaction, and CD-Bonn and AV18 SRC parametrization sets, we predict a heavy neutrino-exchange shell model NME for the $0\nu\beta\beta$ decay of $^{82}$Se
\beq\label{eq12}
M^{0\nu}_N = 158 \pm 29,
\eeq
where the average value and the uncertainty were calculated based on Table \ref{tbl3} pure closure results. The values in Eq. (\ref{eq12}) decrease by less than 0.5\% if instead of $g_A=1.254$ we choose the recent value for the axial-vector constant $g_A=1.269$.

Our analysis suggests that the mixed approximation can be successfully used to obtain the shell-model NME for the $0\nu\beta\beta$ decay of ${}^{76}$Ge,
for which the calculation of the first one hundred intermediate
states for each $J^\pi$ is very challenging but still doable.
Looking at the insert in Fig. \ref{fig4} we expect to have an
uncertainty of the NME for ${}^{76}$Ge within 1\% if about 100 intermediate states will be used.

We were also able to obtain 
for the first time 
a decomposition of the shell-model NME for
light and heavy neutrino-exchange mechanisms versus the spin
of intermediate states and found that for the light
neutrino-exchange the $J=1$ states provide the largest
contribution.
For the heavy neutrino-exchange NME the higher
$J$ are not suppressed and the distribution in Fig. \ref{fig6}
is more or less uniform.

For the future it would be also
interesting to go beyond the closure approximation for the NME of
other isotopes, such as ${}^{76}$Ge, and for other
 mechanisms that could contribute  to the
$0\nu\beta\beta$ decay rates \cite{ves12,prc13,prl13}.

\vspace{0.4cm}

R.A.S. is grateful to V. Zelevinsky for constructive
discussions. Support from  the NUCLEI SciDAC Collaboration under
U.S. Department of Energy Grant No. DE-SC0008529 is acknowledged.
M.H. and B.A.B. also acknowledge U.S. NSF Grant No. PHY-1068217.


\begin{thebibliography} {99}


\bibitem{ves12} J.D. Vergados, H. Ejiri, and F. Simkovic, { Rep. Prog. Phys.} {\bf 75}, 106301 (2012).

\bibitem{jermp}
F.T. Avignone, S.R. Elliott, and J. Engel, Rev. Mod. Phys. {\bf 80}, 481 (2008).

\bibitem{tomoda} T. Tomoda, { Rep. Prog. Phys.} {\bf 54}, 53 (1991).


\bibitem{prc13} M. Horoi, {Phys. Rev. C} {\bf 87}, 014320 (2013).

\bibitem{prl100} E. Caurier, J. Menendez, F. Nowacki, and A. Poves, Phys. Rev. Lett. {\bf 100}, 052503 (2008).

\bibitem{iba-2} J. Barea and F. Iachello, Phys. Rev. C {\bf 79}, 044301 (2009).

\bibitem{iba-prl} J. Barea, J. Kotila, and F. Iachello, Phys. Rev. Lett. {\bf 109}, 042501 (2012).

\bibitem{gcm} T.R. Rodriguez and G. Martinez-Pinedo, Phys. Rev. Lett. {\bf 105}, 252503 (2010).

\bibitem{phfb} P.K. Rath, R. Chandra, K. Chaturvedi, P.K. Raina, and J.G. Hirsch, Phys. Rev. C {\bf 82}, 064310 (2010).

\bibitem{qrpa-J} F. Simkovic, A. Faessler, V. Rodin, P. Vogel, and J. Engel,
Phys. Rev. C {\bf 77}, 045503 (2008).

\bibitem{simvo11} F. Simkovic, R. Hodak, A. Faessler, and P. Vogel, Phys. Rev. C{\bf 83}, 015502 (2011).

\bibitem{prc10} M. Horoi and S. Stoica, {Phys. Rev. C} {\bf 81}, 024321 (2010).

\bibitem{supernemo} A.S. Barabash, arXiv:1112.1784.

\bibitem{sh13} R.A.~Senkov and M.~Horoi, Phys. Rev. C {\bf 88}, 064312 (2013).

\bibitem{jun45} M. Honma, T. Otsuka, T. Mizusaki, and M. Hjorth-Jensen,
Phys. Rev. C {\bf 80}, 064323 (2009).

\bibitem{prl13} M. Horoi and B.A. Brown, { Phys. Rev. Lett.} {\bf 110}, 222502 (2013).

\bibitem{kipf12} J. Kotila and F. Iachello, {Phys. Rev. C} {\bf 85}, 034316 (2012).

\bibitem{barabash10} A.S. Barabash, Phys. Rev. C {\bf 81}, 035501 (2010).

\bibitem{2njun45} E. Caurier, F. Nowacki, and A. Poves, Phys. Lett. B {\bf 711}, 62 (2012).

\bibitem{nushellx} NuShellX@MSU, B.A.~Brown, W.D.M. Rae, E. McDonald, and M. Horoi,\\ http://www.nscl.msu.edu/\~{}brown/resources/resources.html

\bibitem{hpcc} https://icer.msu.edu/hpcc

\bibitem{menendez-sen} J. Menendez, talk presented at the INT Program ``Nuclei and Fundamental Symmetries," INT Seattle, August 5--30, 2013.

\bibitem{npa818} J. Menendez, A. Poves, E. Caurier, and F. Nowacki,
Nucl. Phys. A {\bf 818}, 139 (2009).

\bibitem{npa793} V.A. Rodin, A. Faessler, F. {\v S}imkovic, and P. Vogel,
Nucl. Phys. A {\bf 793}, 213 (2007).

\bibitem{qrpa-tbc} F. {\v S}imkovic, A. Faessler, H. M{\"u}ther, V. Rodin, 
and M. Stauf, Phys. Rev. C {\bf 79}, 055501 (2009).

\bibitem{suh07} M. Kortelainen and J. Suhonen, Phys. Rev. C {\bf 75}, 051303(R) (2007).

\end{thebibliography}
\end{document}